# Precise calibration method for triaxial magnetometers not requiring Earth's field compensation

Ales Zikmund, Michal Janosek
Dept. of Measurement, Faculty of Elec. Eng.
Czech Technical University in Prague
Technicka 2, Praha, Czech Republic

Michal Ulvr, Josef Kupec
Dept. of Electromagnetic quantities, Laboratory of Fundamental metrology, Czech Metrology Institute
V Botanice 4, 150 72 Prague, Czech Republic

*Abstract* – A calibration procedure for calibrations of triaxial magnetometers is presented. The procedure uses a triaxial Helmholtz coil system and an Overhauser scalar magnetometer and is performed in the Earth's field range. The triaxial coils are firstly calibrated with the Overhauser magnetometer and subsequently a triaxial magnetometer calibration is performed. As opposed to other calibration approaches, neither Earth's field nulling system nor movements of the magnetometer are needed. A real calibration test was carried out - the extended calibration uncertainty was better than 430 ppm in sensitivity and 0.06 degrees in orthogonality.

I. INTRODUCTION

A standard way to calibrate triaxial magnetometers is using a compensating system which cancels out any magnetic disturbances [1, 2]. These systems usually utilize 2-m and larger triaxial coils with high homogeneity (Braunbek configuration) and the field compensation is open-loop: it typically uses a triaxial magnetometer placed far away from the building. This triaxial magnetometer is either a standalone where its reading influences the compensating currents, or is placed in another smaller triaxial coil system which runs in a local closed-loop system (maintaining zero field): the compensating current is then shared with the coils at the calibrating facility [3]. This approach has many drawbacks. Mainly, the remote triaxial magnetometer and also the coil system have to be aligned and calibrated precisely to establish stable compensation. Additionally, the zero magnetic field value is not checked during the calibration and this residual field affects the calibration precision.

The alternative closed-loop systems rely on a zero-detector inside the feedback loop which is placed in the cancellation coils, i.e. Billingsley APEX-CS. Ultimate precision is allowed by using atomic magnetometers in the feedback loop [4, 5]. This principle achieves more precise results but is affected by the influence of the zero detector on the calibrated sensor. Both sensors cannot be ideally placed in the center of the coils where the magnetic field is homogeneous, and so their mutual position can increase uncertainty of the calibration.

Another approach is the scalar calibration [6,8], which provides very good results. Nevertheless, it relies on rotations of the calibrated triaxial magnetometer in the Earth's field, and thus is sensitive not only to disturbances but also to magnetic field gradient, requiring magnetically clean locality which is not easy to find even in sub-urban areas. In order to suppress the effect of magnetic disturbances and diurnal variations of the Earth's magnetic field, the magnetic field should be logged (usually with an Overhauser scalar magnetometer) and its values should be used in the calculations. To achieve an uncertainty below 200 ppm, the residual magnetic field variation during the calibration procedure has to be below 0.5 nT. This is not easy to achieve due to the existing magnetic field gradients between the calibrating site and the place where the magnetic field is logged. Similarly, due to the finite size of the calibrated magnetometer the magnetic field should be free of gradients in the whole sphere covered by the magnetometer rotations. This measurement is also very time-consuming because at least 80 different orientations of the magnetometer are usually measured [6].

## II. MOTIVATION

Our motivation, to create an alternative calibrating procedure, is to be able to perform the calibration in a relatively low-cost facility which would be metrologicaly traceable to a magnetic field density standard resulting in less than 0.05% (500 ppm) calibration uncertainty. Such technique would be applicable to a wide community of users as was defined in the European metrology research project JRP IND 08 'MetMags' [7]. In our case, it is necessary only to monitor the Earth magnetic field variations by using Overhauser magnetometer, no shielded rooms or field-cancellation loops have to be used. The calibration facility can also be placed in areas with magnetic field gradients which would not allow the scalar calibration method. The basic requirement is that the calibrating facility uses well-calibrated triaxial coils which would enable to create magnetic fields of up to 100 μT – this is rather standard requirement and such coil systems are commercially available.

## III. THE CALIBRATION

*A. Calibration site*

The site, property of the Geophysical Institute at Academy of Sciences, Czech Republic, has been used till 1960's as a geomagnetic observatory, however, due to building of DC railways and expanding city borders it was later converted for paleomagnetic experiments [9]. Although the site is 4-km away from the nearest subway station and 3-km away from DC electrified railway, the traffic-related noise was observed with disturbances increasing up to 10 nT pp and even the resuming metro operation was identified (see Fig. 1). It is clear that during calibration, the magnetic field should be either compensated or monitored even though averaging might improve the situation but substantially prolonging the calibration time (the time span in Fig. 1 is .1 ¾ hours).

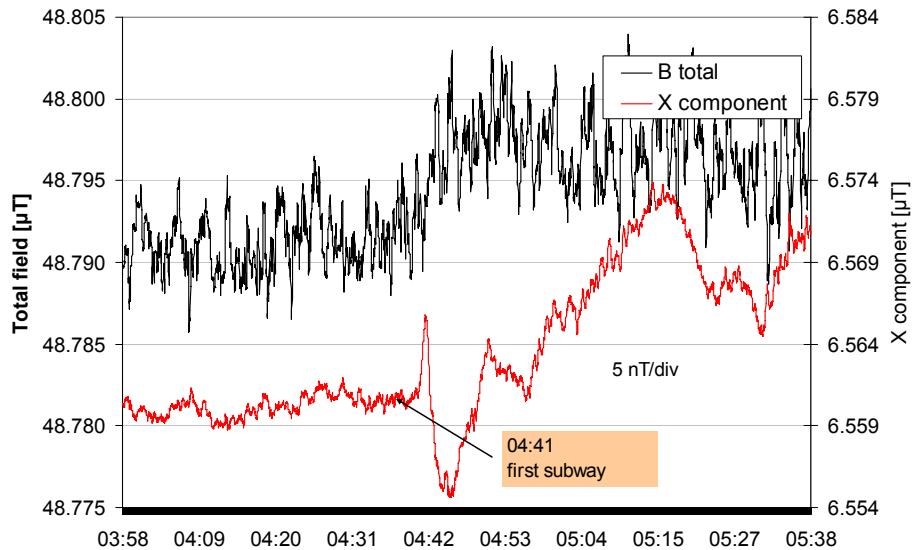

Fig. 1– Noise measurement during traffic strike. The first ongoing subway causes 10 nT p-p noise in horizontal component (approx. E-W).

The coils used for calibration are the commercially available triaxial coils HELM-3 of Billingsley Aerospace & Defense. The coil system consists of three squared Helmholtz coils whose dimension is around 1 meter. The nominal coil constant as provided by the supplier is 100 µT/A (the coils were originally supposed to be used in a feedback system where coil constants do not have to be precisely known). Orthogonality error of the coils is better than 0.1 degree for all three axes. The magnetic field uniformity is declared as 0.3 % in a centrally located 20 cm sphere. For our purpose, however, the coil constants needed to be calibrated.

The current to the coils was supplied by a custom-built current source to overcome drifts of the coil resistance during calibration caused by self-heating and also ambient temperature. The current source is based on 16-bit digital to analog converters and a voltage-to-current amplifier. Stability of the custom-built current source was measured as ± 10 µA in one hour while supplying 1 A. To be absolutely independent of the current source stability, current is simultaneously measured using 1-Ω shunts and 6.5-digit voltmeters.

*B. Triaxial coils calibrations*

The procedure uses a scalar Overhauser magnetometer to calibrate the coil constants and the angular alignment of the triaxial Helmholtz coils – the details are described in [10] together with uncertainty analysis in [11]. The details of the setup are depicted in Fig 2. The calibration results express the 3 coil sensitivities and their respective angular alignment. The results together with the expressed uncertainty are shown in Table I. The resulting coil constants differ from the nominal values about 10 % because the producer does need to define these parameters precisely due to using the coils in the feedback system.

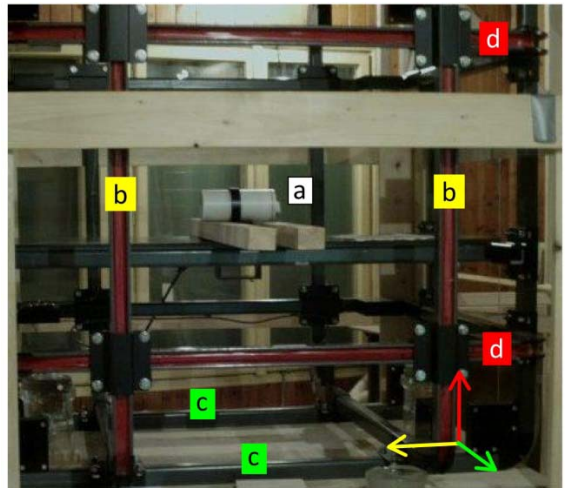

Fig. 2 - a) The Overhauser magnetometer sensor placed in the triaxial coil system during calibration – a central position assures low field gradient, b) the coil pair oriented N-S, c) the coil pair oriented E-W d) the coil pair oriented vertically - from [12].

TABLE I. The parameters of the triaxial system

| Axis | Coil constant [nT/A] | Combined uncertainty [nT/A] |
|---|---|---|
| X – East West | 78787.5 | 16.1 (204 ppm) |
| Y – North South | 76647.5 | 13.4 (175 ppm) |
| Z – Vertical | 83016.4 | 23.7 (286 ppm) |
| **Axes** | **Alignment angle** | **Combined uncertainty** |

|       | [deg] | [deg] |
|-------|-------|-------|
| X – Y | 89.98 | 0.04  |
| Y – Z | 90.01 | 0.04  |
| X – Z | 89.97 | 0.04  |

*C. Calibration against magnetic flux density (MFD) standard*

The aim was to compare the previously calibrated results to a traceable calibration. The calibration method is based on a direct comparison with the magnetic flux density standard, which is basically a 4-section solenoid on a quartz core. The method uses one DC source and is standardized at the Czech Metrology Institute under the procedure code 817-MP-C602 (Fig. 3). The comparison was done for the horizontally located E-W coils only in order to verify the uncertainty of the coil calibration principle [10].

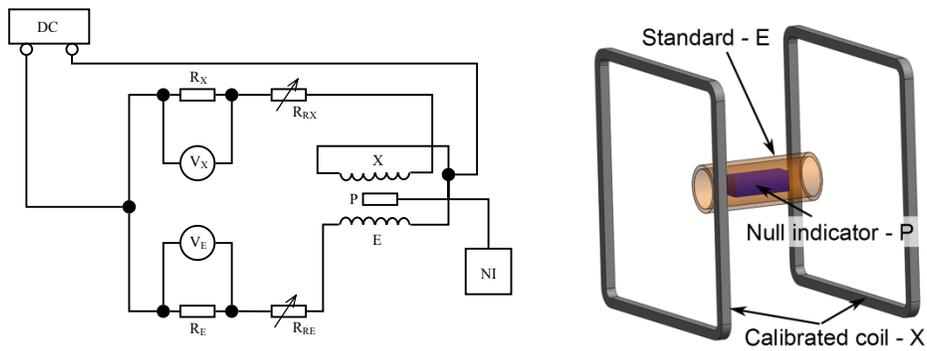

Fig. 3 - Calibration procedure based on comparison with a known MFD standard: a) the electrical circuit b) the real arrangement.

The two coils, the unknown X (the E-W axis of the Helmholtz coils in our case) and the coil standard E were connected in the two branches with variable resistors $R_{RX}$ and $R_{RE}$. The currents in these branches, $I_X$ and $I_E$, were changed in both branches in order to cancel the flux measured in the middle of the standard coil by the means of zero detector P, which was a single-axis fluxgate magnetometer. The current in the two branches is sensed on the resistor standards $R_x$ and $R_E$, respectively. When the zero reading at null indicator (block NI in Fig. 3) was reached, the two currents were recorded and the coil constant $K_{BIX}$ has been calculated as:

$$K_{BIX} = \frac{K_{BIE} I_E}{I_X} = K_{BIE} \frac{U_E}{U_X} \frac{R_X}{R_E} \tag{1}$$

where $K_{BIE}$ is the constant of the standard $E$, $U_E$ is voltage drop on the standard resistor $R_E$, and $U_X$ is the voltage drop on the standard resistor $R_X$.

As the voltage source is common for both branches, its instability is suppressed. The resulting uncertainty can be also suppressed by interchanging the sensing resistors $R_X$ and $R_E$.

The highest uncertainty of 30 ppm had the National flux density standard with a value of 598.827 µT/A. The voltmeters, measuring $V_X$ and $V_E$ were of Agilent 34587A type with 8.5 digits of resolution; $R_E$ and $R_X$ were standard resistors of 2 and 5 Ohm value, respectively, with an uncertainty of 5.4 ppm.

The cross-calibration result of the X (E-W) axis was 78795.5 nT/A with an expanded uncertainty of 50-ppm - this agrees well with the calibration results of the X axis shown in Table I.

*D. Triaxial magnetometer calibration principle*

The basic idea of the calibration has been described in [12]. A predefined sequence of currents is performed together with measuring the response of the calibrated triaxial magnetometer. The Earth's magnetic field is not cancelled by the coil system but its scalar value is remotely monitored. All input quantities (coil current, triaxial magnetometer output and the monitored Earth's field scalar value), forming N equations from N calibration steps where the magnetic field in the calibrating coils is changed in its amplitude and direction, are passed to a solver which solves the problem by the Levenberg–Marquardt non-linear optimization according to [12].

The remote scalar magnetometer measures the Earth's field and should be in a distance so that the influence of the coil system would be negligible. For our experiment, we supposed that the maximum excited magnetic field is around 100 µT, then according to the magnetic field of a dipole source which falls with $1/r^3$ the scalar Overhauser magnetometer had to be placed at least 40 meters away to suppress the coil system influence down to 1.5 nT.

Based on an experience of [12], it was more suitable to approximately align the axes of the calibrated magnetometer with the respective calibrating coil axes because the non-linear solver converged faster. From the optimization described above, the sensitivity and two angles with respect to the orthogonal coil system have been obtained for each axis of the calibrated magnetometer. The resulting parameters are summarized in Table II and depicted in Fig. 4.

TABLE II. The resulting parametrs of the calibration

| Triaxial magnetometer | U | V | W |
|---|---|---|---|
|  |  |  |  |
| Sensitivity | $S_U$ | $S_V$ | $S_W$ |
| Angle to XY coil plane | $\alpha_U$ | $\alpha_V$ | $\alpha_W$ |
| Angle in XY coil plane | $\beta_U$ | $\beta_V$ | $\beta_W$ |

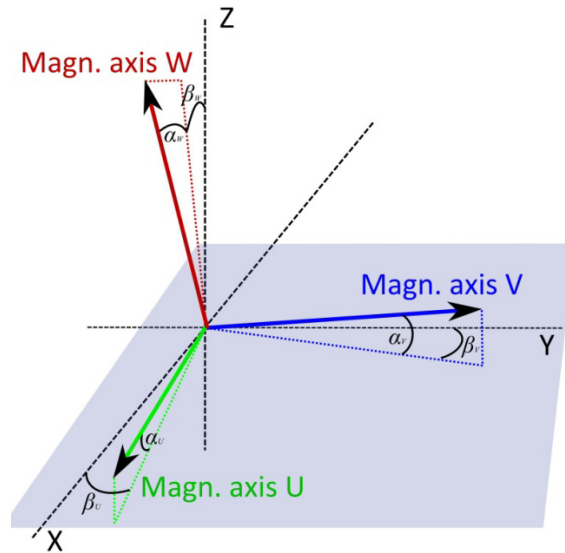

Fig. 4 - The resulting angle parameters.

To compare several calibration principles, the mutual orthogonality angles of the calibrated triaxial magnetometer has been expressed according to [6] by simple calculations:

$$\begin{aligned}\Delta_1 &= \beta_U + \beta_V \\ \Delta_2 &= \alpha_U + \beta_W \\ \Delta_3 &= \alpha_V + \alpha_W\end{aligned} \quad (2)$$

IV. CALIBRATION PROCEDURE UNCERTAINTY.

The triaxial system's coil constants and their orthogonality are known from the previous calibration - see Table III which gives their values and uncertainty. Further errors during the calibration can be caused by the variation of the Earth's magnetic field and by environmental noise inherent to the location. However, both are suppressed to a large level by recording the $B_E$ value with a precise Overhauser magnetometer. The field gradient is subtracted in the measurement and thus the 'residual error' was only taken into account which is the gradient noise; this was measured on site as ± 5 nT.

TABLE III. The uncertainties of the used instruments

| Parameter | Value | Uncertainty |
|---|---|---|
| Overhauser magnetometer [nT] | - | 0.2 nT |
| Coil constants X,Y,Z [nT/A] | 78788, 76648, 83016 | 204, 175, 286 ppm |
| Coil angles XY, YZ, XZ [degree] | 90, 90, 90 | 0.04° |
| Standard resistors values [Ω] | 1.00006, 0.99989, 0.99993 | 30 ppm |
| Voltage measurement | 1 V nominally | 80 ppm |

We used the Monte Carlo method for estimating the influence of all uncertainties of the input variables (Table III) on the resulting parameters as opposed to our initial approach in [12]. The measured parameters, being the input to the optimization method, were deviated with a supposed normal noise distribution. The (B-type) uncertainty was then expressed as a standard deviation of the set of the solver results. The rather high uncertainty of the voltage measurement was caused by the used voltmeter of Agilent 34401 type.

V. EXPERIMENTAL TEST

The test calibration has been carried out in the former geomagnetic observatory site Pruhonice maintained by the Geophysical Institute of Czech Academy of Sciences. A triaxial digital magnetometer that was developed in our laboratory was calibrated in the Helmholtz coil system with calibration parameters in Table III. The ambient magnetic field was first measured with the Overhauser magnetometer in the coils without any excitation and then in the remote spot which was 40 meters far from the coil system - the scalar gradient of 11 nT, which was then subtracted from the data, was supposed to be stable during the whole measurement time.

A predefined current sequence was applied during the calibration [12] (see Fig.5). The sequence contains current steps designed to have an significant influence on the calibrated magnetometer axes in each orthogonal direction and also to keep the magnetic field well in the magnetometer range  The aim was to obtain a response at least 25 µT in each axis of the calibrated triaxial magnetometer.

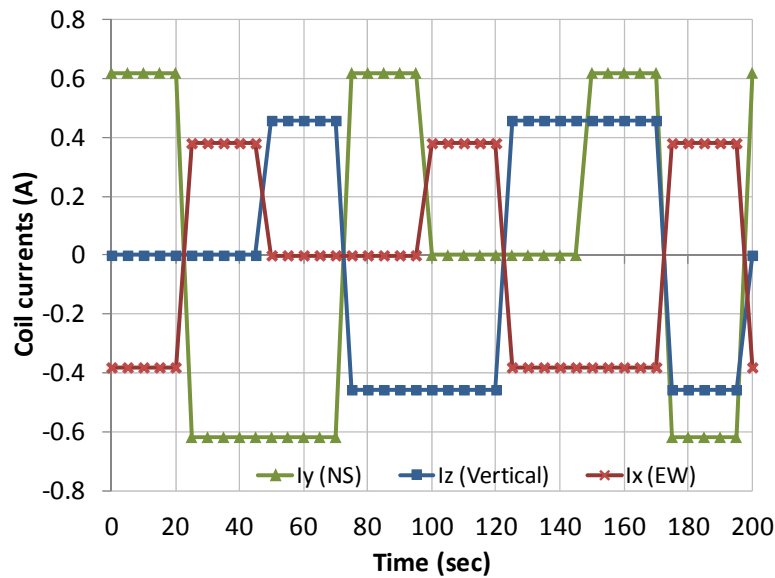

Fig. 5 - The current sequence applied to respective coil axes during calibration of the triaxial magnetometer.

The digital output of the calibrated triaxial magnetometer was recorded during the applied current sequence. In Fig. 6 a time record is shown as a response to the calibrating coils excitation. The magnetic axis orientation is significant because it correlates with the coil excitation. The magnetometer axis W was vertical (coil Z), the axis V was oriented to North-South direction (coil Y) and the magnetometer axis U was approximately aligned with the East-West coil (coil X). The alignment was not ideal due to small cross-field reactions which can be seen in the record, nevertheless, this did not cause a problem for the calibration algorithm.

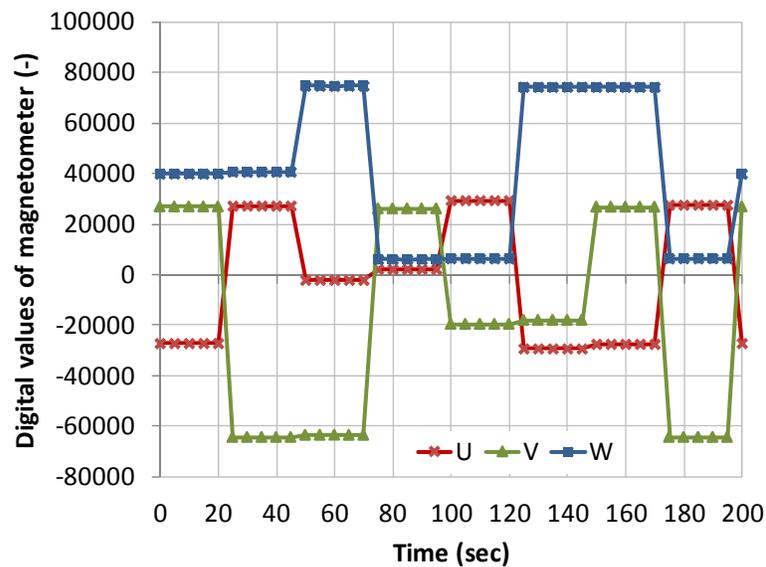

Fig. 6 - The digital triaxial magnetometer response on the reference coil excitation.

The sequence of calibration currents has been repeated 12 times to obtain a minimal statistical set for averaging and calculation of the A-type uncertainty. In the calculation procedure, however, we found higher residuals of the optimization method in some combinations of the current. This was probably caused by vectorial components of ambient magnetic noise which are different - higher residuals were correlated with the situation when the vertical coil was excited. The sensitivities varied maximally of 210 ppm and the angles varied of 0.028 degree which was designated as the A-type uncertainty.

The B-type uncertainties have been established by the Monte Carlo simulation using the parameters from Table III. The input parameters were set up according to the real measured quantities and their values were scattered according to their known uncertainties. The worst B-type gain uncertainty of 110 ppm appeared in the W axis because it was the most affected by magnetic field noise at the location.

The A-type uncertainty was mostly influenced by the noisy magnetic field which is depicted in Fig. 7 (black trace); this is the real data input into the calibration algorithm. The diurnal variation of the Earth's field corresponds to the record of Budkov observatory (Intermagnet designation BDV, red trace). Also the magnetic field gradient variation (or noise), discussed previously, will affect the measurement uncertainty, however, it cannot be measured at the calibration time.

The combined uncertainty was finally calculated as a norm of the two A and B components and the results are given together with the calibrated parameters of the triaxial magnetometer in Table IV. The same triaxial magnetometer calibrated in this work was also calibrated by the scalar calibration, which is a different technique described in [8]. The results are also presented in Table IV to have a comparison. The data agree well – the scalar calibration results are almost within the calibration uncertainty of the developed method. Assuming that also the scalar calibration has a significant uncertainty which is usually expressed as calibration residua [5, 6] -its evaluation is beyond the scope of this paper – we show that our calibration method is at least comparable to the scalar calibration.

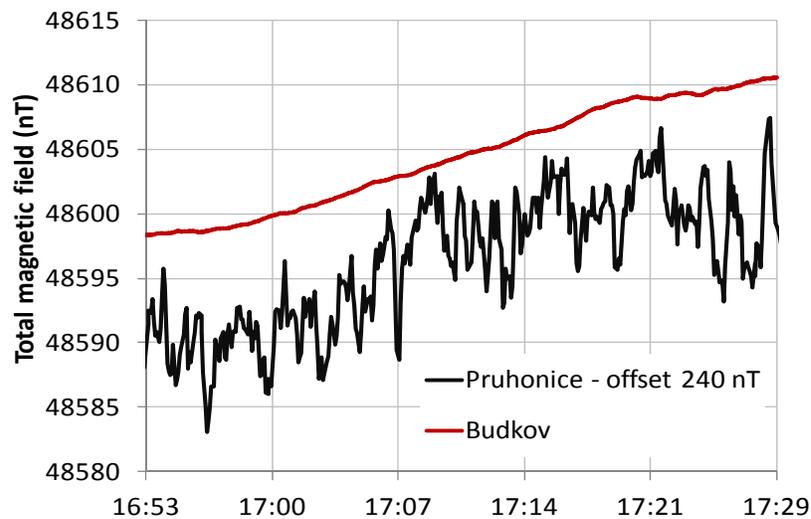

Fig. 7 - The noise of ambient magnetic field during calibration compared to Budkov observatory.

TABLE IV. The uncertainties of the results

|  | Uncertainty B | Uncertainty A | Result with combined | Scalar |
| --- | --- | --- | --- | --- |

|  |  |  | uncertainty (k=2) | method [8] |
|---|---|---|---|---|
| **Offset U ([nT]** | 2 | - | -32.5 ± 4 | -30.5 |
| **Offset V [nT]** | 2 | - | -37.5 ± 4 | -34.4 |
| **Offset W [nT]** | 2 | - | -24.0 ± 4 | -27.6 |
| **Sensitivity U (norm) [-]** | 88 | 132 | 0.9659 ±320 ppm | 0.9666 |
| **Sensitivity V (norm.) [-]** | 68 | 116 | 0.9431 ±270 ppm | 0.9436 |
| **Sensitivity W (norm.) [-]** | 110 | 185 | 0.9022 ±430 ppm | 0.9021 |
| **Angle $\Delta_1$ [°]** | 0.002 | 0.016 | 0.205 ±0.034 | 0.167 |
| **Angle $\Delta_2$ [°]** | 0.001 | 0.024 | 0.531 ±0.048 | 0.603 |
| **Angle $\Delta_3$ [°]** | 0.003 | 0.031 | 0.104 ±0.062 | 0.107 |

## VI. CONCLUSION

The presented calibration procedure is advantageous to the currently used methods because neither an Earth's field cancellation system nor moving a calibrated magnetometer is required to measure the sensitivities and angular misalignments of the respective magnetometer axes. The Earth's magnetic field value was monitored at a distant place with an Overhauser magnetometer and was used in the calibration procedure. The calibration of the used triaxial Helmholtz coils system is performed with the same Overhauser magnetometer as during the triaxial magnetometer calibration, preferably before each calibration, in order to compensate a possible long-term drift of the coil constants. The magnetometer offsets should be measured separately in a magnetic shielding chamber; this is also the way in which the estimation uncertainty is the lowest [13].

From the Monte-Carlo simulations, we have shown that the uncertainty of the calibrated parameters should be less than 260 ppm in sensitivity and 0.02 degrees of arc in orthogonality if the environmental gradient noise is below 5 nT and our measurement precision was gained. A real calibration of a digital triaxial digital magnetometer was done with the proposed procedure. The calibration precision was influenced by gradient noise at the observatory, resulting in the largest combined uncertainty (k=2) of 430 ppm for sensitivity and 0.062 degrees of arc for the orthogonal angle of magnetometer axis.

## ACKNOWLEDGMENT

The research was supported by the European metrology research project 'MetMags', by the grant No TA01010298 of the Technology Agency of the Czech Republic and by the grant No.SGS12/194/OHK3/3T/13 sponsored by Czech Technical University in Prague.

## REFERENCES

[1] H. Kügler, "Simulation of DC Magnetic Environment on Ground," Fourth International Symposium Environmental Testing for Space Programmes, European Space Agency, ESA SP-467, 2001, pp. 263

[2] T. Risbo, P. Brauer, J.M.G. Merayo, O.V. Nielsen, J.R. Petersen, F. Primdahl and I. Richter, "Ørsted pre-flight magnetometer calibration mission", Measurement Science and Technology, Vol. 14, No. 5, 2003, pp. 674-688.

[3] R. Vernier, T. Bonalsky, J. Slavin, "Goddard Space Flight Center Spacecraft Magnetic Test Facility Restoration Project", Nasa Technical Reports, 2004

[4] V.Y. Shifrin, E.B. Alexandrov, T.I .Chikvadze, V.N. Kalabin, N.N. Yakobson, V.N. Khorev and P.G.Park, "Magnetic Flux Density Standard for Geomagnetometers", Metrologia, Vol. 37, No. 3, 2000, pp. 219

[5] K. Weyand, "Maintenance and dissemination of the magnetic field unit at PTB," IEEE Trans. Instrum. Meas., Vol. 50, Issue2, 2001, pp. 470-473

[6] J. M. G. Merayo, P. Brauer, F. Primdahl, J.R. Petersen and O.V. Nielsen, "Scalar calibration of vector magnetometers, " Meas. Sci. Technol., Vol. 11, No. 120, 2000, pp. 120-132

[7] M. Hall et al., "Best practice guide for the generation and measurement of DC magnetic fields in the magnetic field range of 1 nT to 1 mT", NPL report, http://www.npl.co.uk/content/ConPublication/6191, 2014, 32 pp,

[8] V. Petrucha and P. Kaspar, "Calibration of a triaxial fluxgate magnetometer and accelerometer with an automated non-magnetic calibration system," IEEE Sensors 2009 conference, IEEE, 2009, pp.1510-1513

[9] K. Prihoda, M. Krs, B. Pesina, J. Blaha, "MAVACS - a new system of creating a non-magnetic environment for paleomagnetic studies," Cuadernos de Geologia Iberica, Vol. 12, 1989, pp. 223-250.

[10] A. Zikmund, P. Ripka, "Scalar Calibration of the 3-D Coil System," Journal of Electrical Engineering, Vol. 61, No. 7/s, 2010, pp. 39-41.

[11] A. Zikmund, P. Ripka, "Uncertainty Analysis of Calibration of the 3D Coil System", Journal of Electrical Engineering. Vol. 63, 2012, pp. 90-93

[12]  A. Zikmund , M. Janosek  "Calibration procedure for triaxial magnetometers without a compensating system or moving parts" (I2MTC 2014) Proceedings ; 2014, ISBN: 978-1-4673-6385-3, pp. 473-476

[13]  E. Moldovanu, A. Moldovanu, E.D. Diaconu, C. Ioan, B.O. Moldovanu, M. Macoviciuc, "Metrological aspects of the offset evaluation for magnetometric sensors" Sensors and Actuators A: Physical, Volume 59, Number 1, April 1997, pp. 113-118(6)